\documentclass[12pt]{article}
\usepackage{graphicx}
\usepackage{bm,enumerate}

\newcommand\ket[1]{| #1 \rangle}
\newcommand\bra[1]{\langle #1 |}

\newcommand{\beq}{\begin{equation}}
\newcommand{\eeq}{\end{equation}}

\begin{document}
\title{Mathematical Modeling of Physical and Engineering Systems in Quantum Information}

\author{Horace P.~Yuen\\Department of Electrical Engineering and Computer Science, \\
Department of Physics and
Astronomy, \\ Northwestern University, Evanston, IL, 60208, \\
Email: yuen@eecs.northwestern.edu}
\maketitle

\begin{abstract} Several concrete examples in quantum information
are discussed to demonstrate the importance of proper modeling
that relates the mathematical description to real-world
applications. In particular, it is shown that some commonly
accepted conclusions are not adequately supported by their
purported justifications in the logical manner required.
\end{abstract}

\section{Introduction}
This paper describes the major part of my talk at the 2006 QCMC
meeting. (The rest on unconditionally secure quantum bit
commitment are included in my writeup in  this volume for my
poster paper.) It may be viewed as a concrete elaboration of my
points on the role of mathematical modeling and rigor in
real-world applications discussed in [1] and [2]. In the context
of quantum information, such consideration is especially important
and indispensable, if we ever want to develop a true quantum
information technology. There are three main points in the
following that are presented in three sections:

\begin{enumerate}[(i)]
\item Entropy or mutual information that the attacker possesses on
a generated key is not a good quantitative measure of security in
real cryptosystems.

\item Given the entropy criterion, there is no security proof,
unconditional or just limited ones that include all currently
feasible attacks, for any experimental BB84 system.

\item Loss is a major limiting factor on many quantum information systems
that has not been properly dealt with theoretically.
\end{enumerate}

These assertions may appear startling to many, as they are
contrary to the ``accepted opinion'', if I may say, of much of the
quantum information community. However, at this point of writing,
I believe they are incontrovertible truths. I hope this is
substantiated in the following.

\section{Performance Criteria and Security Guarantee}

Let us consider the issue of security measure on the generated key
$K_g$ in a quantum key generation (QKG) system. Eve's Shannon
entropy $H_E(K_g)$, or equivalently her mutual information
$I_E=|K_g| - H_E(K_g)$, is the most commonly used measure. If
Eve's knowledge of $K_g$ is bit by bit, the binary entropy of a
bit is in one-one correspondence with Eve's bit error rate.
However, in general Eve has bit-correlated information on $K_g$,
and we may ask: What is the concrete security guarantee provided
by having $I_E \leq \epsilon$ for a given level $\epsilon$? The
problem arises because $I_E$ or $H_E$ is a theoretical quantity
with no operational meaning automatically attached. In standard
cryptography, this issue does not arise because fresh key
generation is considered impossible [3-5] and was never attempted,
while security of other cryptographic functions is based on
computational complexity.

In ordinary communications, the operational significance of the
entropic quantities is given through the Shannon source and
channel coding theorems, which relate them to the empirical
quantities of data rate and error rate. But what is the
corresponding empirical security guarantee in cryptography? This
issue was \emph{not} addressed by Shannon in his classic
cryptography paper written at about the same time as his classic
information and communication theory papers. It was not addressed
by anybody else since.

It is clear that entropy is just a one-number representation of a
whole probability distribution. All questions of security could be
answered by knowing the complete distribution, which I would like
to call Eve's error profile. Let $p_i = p_E(K_g|Y^ES^E)$, where
$K_g$ is an $n$-bit string, $Y^E$ is Eve's observation, $S^E$ is
her total side information, be ordered $p_1 \geq \cdots \geq p_N,
N = 2^n$. Thus, $p_1$ is Eve's maximum probability of guessing
$K_g$ correctly with her information. As a one-number
representation of $p_i$, $p_1$ is of special importance because it
must be sufficiently small for a meaningful security guarantee.
Even for moderate $|K_g| \sim 10^2 -10^3$, it appears that $p_1
\sim 2^{-20}$ may not be small enough for many applications, while
$p_1 \sim 2^{-10}$ would be a disastrous breach of security.

Generally, if Eve can try $m$ different possible $K_g$ to break
the cryptosystem, the first $m$ $p_i$ are the relevant numbers to
determine any quantitative level of security. For $N$ possible
trials, the trial complexity $C_t = \sum_{i=1}^{N} i\cdot p_i$
which is the average number of trials Eve needs to succeed, is a
meaningful measure of security. The number $p_1$ itself is
operationally meaningful, and is in fact the most suitable measure
if a single number has to be used in lieu of the whole $p_i$.

To illustrate this and the problem of $I_E$, we observe that given
$p_1 \leq 2^{-l}$ for some $l$, we have $C_t \geq (2^l + 1)/2$ and
$I_E \leq n-l$ [3]. In the worst case $p_i$, one has [3],
\begin{equation} \label{worstcaseI}
p_1 \sim 2^{-l} \mathrm{\hspace{1cm} for \hspace{1cm}   } I_E/n
\sim 2^{-l}.
\end{equation}
Thus, if Eve has $10^{-3}$ bit of information per bit of $K_g$, it
is possible that her $p_1 \sim 2^{-10}$. This possibility arises
from the possible correlation between the bits of $K_g$ that is
reflected in Eve's information on the whole $K_g$. The $p_i$ that
gives one deterministic bit of information to Eve out of $|K_g| =
10^3$ in the above example is the best, not the worst case, for
the users. It is not a meaningful procedure to average Eve's $p_1$
or other measure over the possible $p_i$ given a fixed level of
$I_E$, because there is a definite $p_i$ that Eve has for the
given cryptosystem.

Thus, to ensure proper security via $I_E$, one must have $l$
sufficiently small in $I_E/n \sim 2^{-l}$. It is difficult to
ensure exponentially small $I_E$ in an entropic analysis of an
experimental system. In fact, $I_E/n \sim 2^{-10}$ is considered
very good in current experimental BB84 schemes, with $0.1\%$
information leak per bit after error correction and privacy
amplification \cite{tomita06}. But as analyzed above, this does
not rule out the possibility of a disastrous breach of security.
This exponential problem persists if the Kolmogorov distance
$\delta(p,p^0)$ between $p_i$ and the uniform distribution $p^0$
is used in lieu of $I_E$.

It is possible to use privacy amplification algorithms to
guarantee exponentially small $I_E$, but the known result
\cite{bennett95} from Reny\'{i} entropy is always loose by a
factor of 2 in the exponent for bounding $p_1$. In addition,
Reny\'{i} entropy is difficult to deal with quantitatively. There
is no example in quantum cryptography in which it has been
usefully bounded in the finite $n$ case, other than the
i.i.d.~situation which does not cover all the currently feasible
attacks. On the other hand, many theoretical results in
information and communication theory yield directly the
exponential behavior of $p_1$. In this connection, it is important
to observe that $l=\log p_1$ is the true limit on the number of
fresh key bits generated in a QKG or classical key generation
system. Thus, I recommend that $p_1$ be employed as the security
measure in both theoretical and experimental cryptosystem studies.

Here I would like to mention  the following problem of BB84: Since
Eve can break the system completely by a man-in-the-middle attack
if she guesses correctly the message authentication key $K_m$
needed for the public channel, what is the meaning that a much
longer fresh key than $|K_m|$ is still generated?

\section{BB84 Security Proofs}

The assertion I would like to make now is that no complete QKG
protocol has been given with quantified security level that is
proved unconditionally secure in a realistic setting including
inevitable loss and noise. By a \emph{complete protocol} I mean
one which has all the steps specified that can be implemented in a
real system and which goes all the way to yield a final generated
key $K_g$ that has proven security, say $I_E \leq \epsilon$ for a
fixed security level $\epsilon$. Such a complete protocol is
needed by an experimentalist to implement a cryptosystem with
quantified security, granting here that $I_E$ is used as a
security measure. Such a quantified ``secure'' cryptosystem is
what we need to produce to substantiate the claim that we have a
``secure'' QKG system while there is no comparable provenly secure
classical cryptosystem.

This requirement implies that all asymptotic analysis and random
coding existence proofs with no finite code specified are not
sufficient for a real cryptosystem that always has a finite bit
length and that requires explicitly specified protocol steps.
Indeed, even if a code is specified, it is not ``realistic'' when
the decoding cannot be carried out in polynomial time. This is
especially the case in view of the fact, to be shown elsewhere,
that exponential inefficiency can be utilized to generate fresh
key via exponentially small probability of success.

There are as yet only two papers with unconditional security claim
on the finite realistic protocols presented that include the
effect of noise. In \cite{inamori01},  the code is not specified
and it is not clear if a code, especially one with polynomial-time
decoding algorithm, can be found in a typical experimental
parameter region. In \cite{hayashi06}, the final quantitative
result is derived under approximation without rigorous bounds.
Both \cite{inamori01} and \cite{hayashi06} do not include all
possible attacks in a system with transmission loss, as is the
case for every security proof given thus far.

The typical loss of linear attenuation is inevitable and usually
considerable in an optical system. Its effect on the security of
BB84 or Ekert type cryptosystem has never been rigorously
determined, while it is accepted by many in the ``community'' that
it only affects the throughput of a single-photon BB84 system but
not its security. That this has \emph{not} been established by a
proper analysis is readily seen from the fact that the qubit model
is not applicable to the situation where the transmission medium
is lossy. If only the detector is lossy, but Eve is assumed not
able to manipulate it, a good assumption in most cases, the qubit
model holds for transmission. If the line is lossy, and Eve is
assumed capable of introducing an alternative lossless medium,
there are additional attacks she could launch on single-photon
BB84 similar to the case of coherent-state or multi-photon BB84.
Even at the individual attack level, she could launch an
approximate probabilistic cloning attack similar to the
unambiguous state discrimination attack in the multi-photon case.

With probabilistic cloning, it is possible to clone nonorthogonal
linearly independent states with a nonzero, and of course
nonunity, probability \cite{duan98}. Similarly, it may be possible
to approximately $n$-clone any set of states with a nonzero
probability and fidelity larger that that obtainable with unity
probability. This possibility has been explicitly demonstrated
\cite{fiurasek05}. By adjusting the probability of success for a
given loss level, Eve could launch such an attack on single-photon
BB84 without being detected. If the resulting fidelity in a
2-clone is higher, Eve's attack becomes more powerful in the lossy
case.

This possibility has not been analyzed in the literature, although
for individual attacks it can be shown that the fidelity cannot be
increased in this way \cite{nair06}. However, this already shows
that a 3-level model of a qubit in loss is necessary to represent
the physical situation, so that all possible attacks by Eve are
accounted for in a joint attack. Actually, an infinite-dimensional
multimode model should be used. This analysis is currently
lacking. In particular, the use of decoy states \cite{hwang03} in
multi-photon BB84 in loss does not solve the security problem of
such sources in loss, because at best the problem of single-photon
BB84 in loss remains.

Actually, I believe there are several problems in the current
proofs of BB84 security that make the validity of various
arguments quite questionable. They will be addressed elsewhere. A
lot of these problems center around the issue of how one may be
able to make rigorous assertions on a multi-correlated system by
examining just one copy. It also appears to me that only
symmetrized joint attacks are included in the current proofs. It
has not been shown why unsymmetrical attacks, especially adaptive
ones, could not do better. These problems disappear in the case of
individual attacks. However, for such attack, the problem of fully
accounting for the side information that can be exploited in just
collective classical processing is difficult, and many errors on
this issue in the literature can be found \cite{yamazaki07}. On
the other hand, there is no such problem in the KCQ (Keyed
Communication in Quantum Noise) approach \cite{yuen03,yuen07}.

Note added: The issues of Sections 2 and 3 are being addressed by
M.~Hayashi and applied to the experiment of A.~Tomita. The
security and efficiency analysis of BB84 including especially
message authentication will be presented by our group shortly.

\section{Loss in Quantum Metrology and Quantum Computation}
Loss is a major limiting factor on the quantum effects obtainable
in a physical system, which is well-known in quantum optics in the
case of squeezing \cite{yuen03book} and especially superposition
of ``macroscopic states'' \cite{caldeira85,walls85}. Recently, it
has been proposed that the NOON state $|\psi\rangle =
\frac{1}{\sqrt{2}}(\ket{N}\ket{0}+\ket{0}\ket{N})$ for number
state $\ket{N}$ could lead to  improved interferometric
measurements with , e.g., a phase resolution $\Delta\phi \sim 1/N$
instead of the $\sim 1/\sqrt{N}$ obtained with coherent states.
They would find many applications under the heading of ``quantum
metrology''. Actually, squeezed states alone on a single mode
would lead to such improvement without entanglement, which is the
optimum value obtainable for a fixed $N$ \cite{yuen03book}. Also,
the state
\begin{equation} \label{state}
\ket{\phi}=\frac{1}{\sqrt{2}}(\ket{N}\ket{N-1}+\ket{N-1}\ket{N})
\end{equation}
leads to similar improvement \cite{yuen86} as the NOON state
$\ket{\psi}$, and can be more closely generated by optical
parametric processes.

Superposition of macroscopic states is ``supersensitive'' to loss.
I have re-emphasized the significance of this phenomenon in
quantum information \cite{yuen05jap,yuen96}. Consider the state
$\frac{1}{\sqrt{2}}(\ket{n_1}\ket{n_2}+\ket{n_2}\ket{n_1})$ for
number states $\ket{n}$, with $\rho$ the corresponding density
operator. Let $\rho'$ be the incoherent superposition $\rho'=
\frac
{1}{2}(\ket{n_1}\ket{n_2}\bra{n_1}\bra{n_2}+\ket{n_2}\ket{n_1}\bra{n_2}\bra{n_1})$.
If the system is in typical linear loss with transmittance $\eta$,
it is readily computed that the trace distance is \cite{yuen96}
\begin{equation} \label{tracedistance}
\|\rho - \rho'\|_1 = 2\eta^{n_1+n_2}.
\end{equation}
Thus, for large $n_1 + n_2, \|\rho - \rho'\|_1 \sim 2e^{-1}$ and
the system effectively decoheres with the loss of one photon. For
a large $N$ NOON state, a fractional loss of $1/N$ would already
destroy the quantum effect responsible for the $\Delta\phi$
improvement. Furthermore, (3) shows qualitatively that the
entanglement effect responsible for the improvement of any usual
performance criterion is wiped out with a tiny loss. This should
remain true for any other entanglement of macroscopic states. The
coherent state case is also worked out in ref. \cite{yuen96}.

I believe a similar supersensitivity obtains in a long multi-qubit
entanglement for quantum computation, which cannot be removed by
fault-tolerant quantum computing or ``quantum leak plumbing''. The
reason is that the terms in a long superposition of many qubits
also contain many quanta, which would become supersensitive in the
presence of loss similar to the NOON state. The situation cannot
be rectified by fault-tolerant qubits which are themselves lossy.
Also, quantum leak plumbing disturbs the system in an
unpredictable way even if no leak is found. To my knowledge, this
whole issue has not been properly treated theoretically in the
literature. While linear loss is significant in all current
experimental quantum computation schemes, there are many other
theoretical schemes in which such loss can be made negligible.
However, the moral I would like to draw here is that we should
incorporate all the small but perhaps ultimately significant
perturbations in the theoretical study of quantum information
systems, and one should not believe that a system would do what it
is designed for without such perturbations and small details fully
taken into account in the system model.

\section{Conclusion}
There is currently a huge divide between theory and experiment on
quantum information systems, even just on a small scale. I believe
this arises also from inadequate modeling of the system as in the
large scale case discussed above. In cryptography, there is the
further complication that security guarantee has to be obtained
with mathematical rigor, assuming the model is complete and
correct. It is possible to show that a cryptosystem is insecure by
an experiment or a simulation, but it is not possible to
\emph{prove} a cryptosystem secure by such means or by other
qualitative reasoning. This point I made in \cite{yuen97} comes in
full force for security guarantee. We should be extra careful in
our modeling and proofs of quantum cryptographic systems. Finally,
there is the question whether any useful concrete system can be
built for a realistic application if it is so model-sensitive as
in the BB84 case, an issue we have not discussed but is widely
known.

\section{Acknowledgement}

I would like to thank Eric Corndorf, Won-Young Hwang, Max
Raginsky, and especially Ranjith Nair on many useful discussions
on the topics of this paper, which was supported by DARPA under
grant F 30602-01-2-0528 and AFOSR under grant F A 9550-06-1-0452.

\end{document}